\documentstyle[11pt,aaspp4,psfig]{article}  

\newcounter{saveeqn}
\newcommand{\alpheqn}{\setcounter{saveeqn}{\value{equation}}%
\stepcounter{saveeqn}\setcounter{equation}{0}%
\renewcommand{\theequation}{\mbox{\arabic{saveeqn}-\alph{equation}}}}

\begin{document}

\title{Absolute Magnitude Calibration for the W~UMa-type Systems
Based on {\sc Hipparcos} 
Data\footnote{Based on data from the ESA Hipparcos astrometry satellite}}

\author{{\sc Slavek M. Rucinski\altaffilmark{2}}\\
\rm Electronic-mail: {\it rucinski@cfht.hawaii.edu\/}}
\affil{Canada -- France -- Hawaii Telescope Co.\\
P.O.Box 1597, Kamuela, HI 96743}

\and

\author{\sc Hilmar W. Duerbeck\altaffilmark{3}\\
\rm Electronic-mail: {\it hilmar@cygnus.uni-muenster.de}}
\affil{Space Telescope Science Institute, Baltimore, MD 21218}

\altaffiltext{2}{Affiliated with
York University and University of Toronto, Canada}
\altaffiltext{3}{Affiliated with M\"unster University, Germany. 
Permanent address: PF 1268, 54543 Daun, Germany.}

\centerline{\today}


\begin{abstract}
Hipparcos parallax data for 40 contact binary
stars of the W~UMa-type (with $\epsilon M_V < 0.5$)
are used to derive a new, $(B-V)$-based absolute-magnitude 
calibration of the form $M_V = M_V\,(\log P,\,B-V)$. The calibration 
covers the ranges $0.26 < (B-V)_0 < 1.14$, $0.24 < P < 1.15$ day, 
and $1.4 < M_V < 6.1$; it is based on a solution weighted by relative
errors in the parallaxes (2.7\% to 24\%). Previous calibrations
have not been based on such a wide period and color space, and
while they have been able to
predict  $M_V$ with sufficient accuracy for systems
closely following the well-known period-color relation, the
new calibration should be able to give also good predictions for 
more exotic ``outlying'' contact binary systems.
The main limitations of this calibration are the inadequate 
quality of the ground-based photometric data, and the restriction
to the $(B-V)$ index, which is more sensitive to 
metallicity effects than the $(V-I)$ index; metallicities are,
however, basically unknown for the local W~UMa-type systems.
\end{abstract}

\section{INTRODUCTION}
\label{intro}

Contact binary stars of the W~UMa-type are -- externally -- 
simple structures. They are single objects with two mass centers and
surfaces described by common equipotentials so that
fewer parameters are needed to describe them than is the case for
detached binaries: In place of two radii, one value of potential is
needed, and in place of two surface temperatures, apparently one for
the common surface suffices (Rucinski \markcite{ruc85}1985,
Rucinski \markcite{ruc93}1993). The poorly understood 
complexity of their internal structure is hidden from our view.
The external simplicity and the observational property of the W~UMa-type 
systems to appear close to the main sequence were the underlying 
reasons for establishing an absolute-magnitude calibration
(Rucinski \markcite{ruc94}1994 = CAL1) in terms of 
two observational quantities, the orbital period $P$ 
and an intrinsic color, e.g.\ $(B-V)$ 
or $(V-I)$. Both quantities are correlated through the 
combined effects of similar geometry, Kepler's third law and
main-sequence relationships, although the correlation is 
not perfect because of the evolutionary spread in sizes 
within the main sequence.

The existence of a period -- color -- luminosity relation is somewhat
reminiscent of that for pulsating stars. This should not be surprising if
one realizes that the orbital periods for similar contact
structures are in a simple relation to dynamical (or free-fall) time
scales and that the radiating areas simply scale with the length of
the period. 
A calibration of the form $M_V = a_P\, \log P + a_C\, C + const$
(where $C$ is for any color index) can be also considered as an 
attempt to shift the stress from the color-derived
temperature dependence of the normal main sequence (which presents 
operational difficulties, as it is affected by interstellar
reddening) to a period dependence controlled by Kepler's third law.
For the absolute magnitude calibration and derivation, the period can
be considered essentially error-free, because it can be precisely 
determined by studying accumulated deviations of times of minimum
light from a given ephemeris. 

As explained in CAL1, no mass-luminosity relation
is involved in the calibration, just the simple geometrical relations.
The most important omission consists of 
the neglect of the mass-ratio ($q$) term. 
The mass-ratio for newly discovered or faint systems
is usually unknown as its determination requires spectroscopic 
observations at large telescopes, which are difficult to obtain for 
these typically faint and very short-period systems. 
Fortunately, the $q$-term possibly correlates with, 
or can be absorbed by, the two other terms. The other 
limitation of any luminosity calibration for contact binaries is 
that it is likely affected by the presence of spots on such
very active stars. Therefore, the calibration applies to
a somewhat poorly defined mean maximum brightness, where
the averaging is over a spot-induced variability of poorly known temporal
characteristics. 

The first calibration (CAL1) was based on 18 systems, 
mostly members of high galactic-latitude open clusters 
(to avoid Milky Way interlopers in more typical open clusters), 
and of visual binaries, as well as on 3 nearby systems with known 
trigonometric parallaxes, 44i~Boo~B, VW~Cep and $\epsilon$~CrA. 
Even for these three systems, the data were somewhat uncertain: 
The color of 44i~Boo~B had not been measured with any degree of
accuracy due to the light of the close visual companion, 
and large discrepancies existed between published parallax values 
for $\epsilon$~CrA. From the beginning, there was a need for reliable 
trigonometric parallaxes for a large number of systems. 

CAL1 concentrated on the $M_V = M_V (\log P,\,B-V)$ relation. Further 
work related to an approximate evaluation of the metal abundance 
effects, based on systems in very metal-poor 
clusters (Rucinski \markcite{ruc95}1995 = CAL2), and to a calibration 
of the relation in the form $M_I = M_I (\log P,\, V-I)$, to study the 
galactic distribution of systems discovered in
the OGLE microlensing search (Rucinski \markcite{ruc97a}1997a 
= CAL3).  In summary, the established relations were:
\begin{eqnarray}
M_V & = & - 2.38\, \log P + 4.26 \,(B-V)_0 + 0.28 - 
0.3\,\rm [Fe/H] \label{eq1}\\
M_V & = & - 4.43\, \log P + 3.63 \,(V-I)_0 - 0.31 - 
0.12\,\rm [Fe/H] \label{eq2}\\
M_I & = & - 4.6 \, \log P + 2.3\,(V-I)_0 - 0.2 - 
0.12\,\rm [Fe/H] \label{eq3}
\end{eqnarray}
where the first two equations combine the results of CAL1 and CAL2,
while the third combines the results of CAL2 and CAL3. 
The colors in Eqs.~\ref{eq2} and \ref{eq3} are de-reddened 
and the $(V-I)$ color is measured in the Kron -- Cousins system. 
The $(V-I)$-based calibrations (Eqs.~\ref{eq2} and \ref{eq3}) 
must be considered preliminary because of the lack of 
directly observed colors.
Due to strong inter-parametric correlations, 
the coefficients in the relations 
have very large and highly non-Gaussian errors (see CAL1 and
below, Section~\ref{calib}). However, the 
absolute magnitudes can be predicted quite accurately. 
As was illustrated through Monte Carlo simulations, 
the preliminary $M_I$ calibration (Eq.~\ref{eq3}) could 
predict the absolute magnitudes to one-sigma (68.3\% confidence)
level of $\pm 0.25$ magnitude and to two-sigma (95.4\% confidence)
level of $\pm 0.5$ magnitude (see Figure~3 in 
Rucinski \markcite{ruc96}(1996)). At first, the period range was
only $0.27 < P < 0.59$ day, but was slightly extended to 0.63 day 
in CAL3.
   
The extant versions of the luminosity calibration have been used 
by several authors, mostly for consistency checks on 
the membership of individual newly discovered systems 
in various clusters (Edmonds et al.\ \markcite{edm96}1996, 
Kaluzny et al.\ \markcite{kal96a}1996a,
Kaluzny et al.\ \markcite{kal96b}1996b,
Mazur et al.\ \markcite{maz95}1995, 
Rubinstein and Bailyn \markcite{rub96}1996, and 
Yan and Mateo \markcite{yan94}1994).
A somewhat more ambitious  
application was its use for distance {\it determinations\/} 
within the ``pencil-beam'' search volume of the 
micro-lensing project OGLE (CAL3). 

Although the predictive power 
for individual cases is relatively low (conservatively
estimated to be more accurate than 0.5 magnitude, taking into account the
possibility of systematic errors), the various forms of the 
the calibration have a definite potential. Of particular importance is
the high frequency of occurrence of the contact binaries:
Disregarding the difference in population characteristics, 
the contact binaries are some 24,000 times more common 
than RR Lyr stars (CAL3) in the solar neighborhood. 
Thus, one can imagine their application to 
detailed galactic structure studies, in place
or in addition to those based on RR~Lyr stars. In addition, 
an extension of the calibration towards early-type 
contact systems may enhance their usefulness
for more distant stellar systems. The calibrations may be also useful
in obtaining a better estimate of the spatial density of contact systems. 
The apparent density derived from the OGLE sample (CAL3) of about
one such system per 250 -- 300 main sequence stars of similar spectral
types (F--K) is about 3 to 4 times higher than the density estimated
from a sample of systems in the solar neighborhood (Duerbeck
\markcite{due84}1984). The discrepancy may be due
to incompleteness of the contact binary sample
below the brightness level of about $V = 8$. The
complete sample of the F--K stars in the Hipparcos Input
Catalogue (Turon et al.\ \markcite{tur94}1994) 
contains 128, 497, 1620 and 4561 stars of luminosity classes
IV and V within the range of occurrence of the W~UMa-type systems, $0.4 <
(B-V) < 1.2$ to magnitude limits $V = 5, 6, 7, 8$. To the same brightness
limits, there are 3, 3, 6, and 14 contact binaries giving the inverse
relative frequencies 43, 166, 270, 326. Thus, the frequency derived
from the OGLE data is approximately confirmed, but with a 
large statistical uncertainty and with a trend indicating that
discovery selection effects become progressively stronger for fainter
systems. Once a larger sample of such faint (i.e.\ not accessible to
direct parallax measurements) contact
systems is identified, the calibration can be used to define a
volume-limited subsample for an unbiased local density estimate.

Parallax data from the {\sc Hipparcos} satellite (ESA
\markcite{esa97}1997) have already
been used to derive a new absolute-magnitude calibration for the
W~UMa-type systems (Rucinski and Duerbeck
\markcite {ruc97c}1997 = CAL4).
The main motivation was the obvious need of having
a solid check on the previous calibrations which had been
based on inhomogeneous data from various sources. A sample 
defined by the errors in $M_V$ smaller than 0.25 magnitude 
($\epsilon M_V = 2.17 \frac{\epsilon\pi}{\pi} < 0.25$) 
consisted of 20 systems, but the system with the
largest and most precisely measured parallax, 44i~Boo~B,
still does not have a reliably measured color and had to be
omitted. Only 2 systems were common to CAL1 and CAL4. The calibration
directly comparable to CAL1 is CAL4a:
\begin{eqnarray}
M_V & = & - 4.30\,\log P + 3.28\,(B-V)_0 + 0.04 \label{eq4a}
\end{eqnarray} 
CAL4a turned out to show a small standard error ($\sigma = 0.17$)
and, despite its somewhat differing coefficients from those of CAL1,  
to give encouragingly similar predictions on $M_V$ as compared to 
the previous calibrations. Monte Carlo simulations 
showed that, as far as random errors are considered, 
absolute magnitudes of contact binaries can be predicted with an accuracy of 
about 0.1 magnitude, which is surprisingly good  in view of 
the omission of the mass-ratio term and of the possibility of star spots.
However, deviations of individual systems from the ``fundamental plane'' 
of the calibration indicated the possibility of 
systematic rather than random errors, and the problem of the 
applicability of the calibration over wider ranges of periods and
colors remained. CAL4 was established within the period range 
$0.27 < P < 0.65$ day, with most calibrating systems clustering around
the main period--color relationship. The experimental addition of 
the long-period, red system V371~Cep (V5 in NGC~188)
led to very different calibration coefficients. Since the two
quantities of the calibration, period and color, are strongly 
(but not perfectly) correlated, the calibration 
is extremely sensitive to the inclusion of
severely spotted or non-contact systems mimicking genuine W~UMa-type 
systems. Realizing these limitations of CAL4, we
consider in the present paper a larger sample, comprising also
systems with poorer parallax data, with a hope that such a sample 
would cover a wider range of periods and colors, and thus lead to
better averaging of individual peculiarities than in the case
of the 19 systems used to establish CAL4.

\section{Hipparcos and auxiliary data}
\label{data}

Our present sample consists of 40 systems with parallax errors
implying uncertainties $\epsilon M_V < 0.5$ magnitude. Most systems
are well-known, bright W~UMa-type binaries. While details on the
adopted photometric data for the sample systems are given in the next
section and in the extensive Appendix, 
several systems included in or excluded from the sample require 
a special mention: 
\begin{enumerate}
\item [44i~Boo] The color data for the contact binary which is the
fainter component (B) of a close visual system are too 
uncertain to warrant its inclusion in the sample. 44i~Boo~B is
the nearest contact binary with a parallax of $78.4
\pm 1.0$ mas (milli-arcsec).
\item [BW~Dra] We included this system although the relatively
uncertain parallax data ($15.3 \pm 5.2$ mas) would formally
exclude it from consideration. This system forms 
a visual binary with another contact system, BV Dra. The
better-quality parallax of BV~Dra was assigned to BW~Dra, too.
\item [AP~Dor] The poorly known contact binary HD~33474
was included in the sample although only a fragmentary light 
curve and a coarse estimate of the period existed until now (Eggen 
1980). The value of the period has been taken form the Hipparcos Catalogue 
(ESA 1997) where it is classified as a W~UMa or possibly a
RR~Lyrae star. However, 
Eggen \markcite{egg80}(1980) argued convincingly that AP~Dor is a contact 
binary system. Its absolute magnitude also contradicts
the RR~Lyrae classification.
\item [BV~Eri] This binary was initially included in the list of
contact binaries to be studied with Hipparcos (HIC~18080), but it 
has been excluded from our present study since the photometric 
analysis of  Baade et al.\ \markcite{baa83}(1983) has shown that 
it is a close
detached binary.
\item[UZ~Oct] This contact system has period slightly
longer than one day and
formally does not belong to the class of W~UMa-type systems. We
included it in our calibration. It contributes little to the
final determination, but its deviation from the calibration
established for short-period systems may give us indications of the
expected trends for long-period systems.
\item [ER~Ori] According to the photometric and spectroscopic
analysis, ER~Ori should have a trigonometric parallax between
5 and 7 mas, which would probably have lead to the inclusion 
into our sample. The Hipparcos parallax is, however, $-6.8 \pm 1.4$ 
mas. Since ER~Ori is a triple
system with a third component at a distance of $0''\!.1$, found by
speckle observations (Goecking et al.\ \markcite{goe94}1994), 
it is likely that the orbital motion 
around the common center of mass has resulted in an unrealistic 
parallax value.
\end{enumerate}

In addition to the parallaxes and the periods, the input data 
for the calibration consist of the observed maximum magnitudes
in $V$, the $(B-V)$ colors at maximum brightness, and the reddening 
values $E(B-V)$. No measured values of $(V-I)$ were available 
for most systems, thus no separate $M_V=M_V(\log P,\,V-I)$ calibration 
was attempted (in most cases, the  $(V-I)$ data listed in the final Hipparcos 
Catalogue have not been measured, but were derived from $(B-V)$). 
This color index has become a standard one in CCD-based 
studies of stellar clusters and micro-lensing searches, and it 
is unfortunate that we presently do not have a good database to carry out
a calibration.

The $V$ magnitude and $(B-V)$ color index at maximum light 
were usually taken from the literature, but sometimes had to be
inferred from Hipparcos magnitudes or from Str\"omgren photometry. 
Details on all systems are given in the Appendix to permit a quality
assessment in each case.
The adopted values of the data used to establish the new calibration 
are listed in Table~\ref{tab1}. It should be stressed that some of these 
input data have gone through a much more careful scrutiny and thus
supersede those adopted in CAL4. Although our main goal was to
establish reliable values of the magnitudes and colors at light maxima,
both Table~\ref{tab1} and the Appendix update and extend a 
very useful compilation of parameters of contact systems by
Maceroni and Van't Veer \markcite{mac96}(1996), which focussed on
contact systems with modern photometric solutions. Our Table includes
14 systems not listed by them -- in most cases because no detailed
analysis has been carried out until now.

As indicated, the available $V_{\rm max}$ data for some systems 
were poorly known or considered not trustworthy. 
In such cases, we gave a higher weight to independent estimates of 
$V_{\rm max}$, based on the Hipparcos data for the upper 5\% of
the brightness (the 95-percentile levels), $Hp_{\rm max}$. Since the 
$Hp$ photometric system slightly differs from $V$,
a transformation to $V$ magnitudes was necessary. In CAL4, we used a
$Hp$ to $V$ transformation involving a small $(V-I)$-term which was
estimated from the adopted $(B-V)$ via the
color -- color calibrations of Bessell \markcite{bes79} (1979).
This two-step process has been simplified here by 
considering differences $Hp-V$ directly 
as a function of $(B-V)$. Reliable values of
$Hp-V$ at maximum light for 30 of our systems were used to
establish two- and one-parameter relationships, $Hp - V = 0.17
(B-V) + 0.03$ and $Hp - V = 0.22 (B-V)$. We adopted the latter, noting
that the intercept should be zero by definition, but that in the
present case it may be non-zero because of the averaging at the
95-percentile level in $Hp$.

If no $(B-V)$ index was available, it could sometimes be calculated 
from a measured $(b-y)$ index, using relations
established by Bessell \markcite{bes79}(1979). Some $(B-V)$ indices were
taken from an unpublished photometric survey of southern contact binaries 
(Duerbeck \markcite{due97}1997). In all instances, the data were
compared with the mean data from the Tycho experiment tabulated in 
the Hipparcos Catalogue. 

The reddening $E(B-V)$, although a relatively small quantity 
for the nearby systems considered here, can affect the calibration
in a systematic way. It tended to be over-estimated in 
previous studies (Rucinski and Kaluzny \markcite{ruc81}1981, 
Rucinski \markcite{ruc83}1983) through the use of
general distance and galactic-latitude scaling laws, which 
may be invalid for small distances. In fact, the mean 
value of  the extinction of  $A_V = 1.8$ mag/kpc 
(Whittet \markcite{whi92}1992) applies only 
to situations of good averaging of many individual 
interstellar clouds at large distances. For moderate 
distances below 100 pc, and well within the galactic
disk, the interstellar medium shows 
strong inhomogeneities, with some directions almost devoid of any
absorption. The novel approach, adopted in CAL4 and in the present paper, 
uses values of neutral hydrogen column density, $N_{\rm HI}$,
to estimate $E(B-V)$. A large database of  
determinations for $N_{\rm HI}$ is available as an Internet
hydrogen column density search tool from the Center 
for Extreme Ultraviolet Astrophysics in Berkeley, 
California\footnote{http:$//$www.cea.berkeley.edu/$\sim$science/html/sci\_archive\_tools\_tools.html.}; 
it is based on the data in Fruscione et al.\ \markcite{fru94}(1994). 
On the basis of the studies by 
Bohlin et al.\ \markcite{boh78}(1978) and 
Tinbergen \markcite{tin82}(1982), the adopted relation between $E(B-V)$ 
and $N_{\rm HI}$ (cm$^{-2}$) was: 
$E(B-V) \simeq 1.7 \times 10^{-22} N_{\rm HI}$. 
A newer study, based on X-ray data (Predehl and Schmitt \markcite{pre95}
1995), gives an almost identical relation with the coefficient
of $1.8 \times 10^{-22}$, a difference which can be entirely 
neglected for the small values of $E(B-V)$ encountered in here.
The values of $E(B-V)$ in Table~\ref{tab1} derived in the above way
carry uncertainties of about 0.01 -- 0.02 magnitude. On the average, they
are much smaller than previously assumed for the field W~UMa systems,
but remain uncertain due to interpolation of the values of $N_{\rm
HI}$ over an irregular grid of stars, with relatively
large spacings in angular separation and distance.

The absolute magnitudes have been derived from 
$M_V = V + 5\,\log \pi - 3.1\, E(B-V) - 10$, where the parallax
$\pi$ is in milli-arcsec units ($0''\!.001$), and the 
observed magnitude at maximum light is in the Johnson $V$-band.

\placetable{tab1}

\section{The new absolute-magnitude calibration}
\label{calib}

\begin{figure}           
\centerline{\psfig{figure=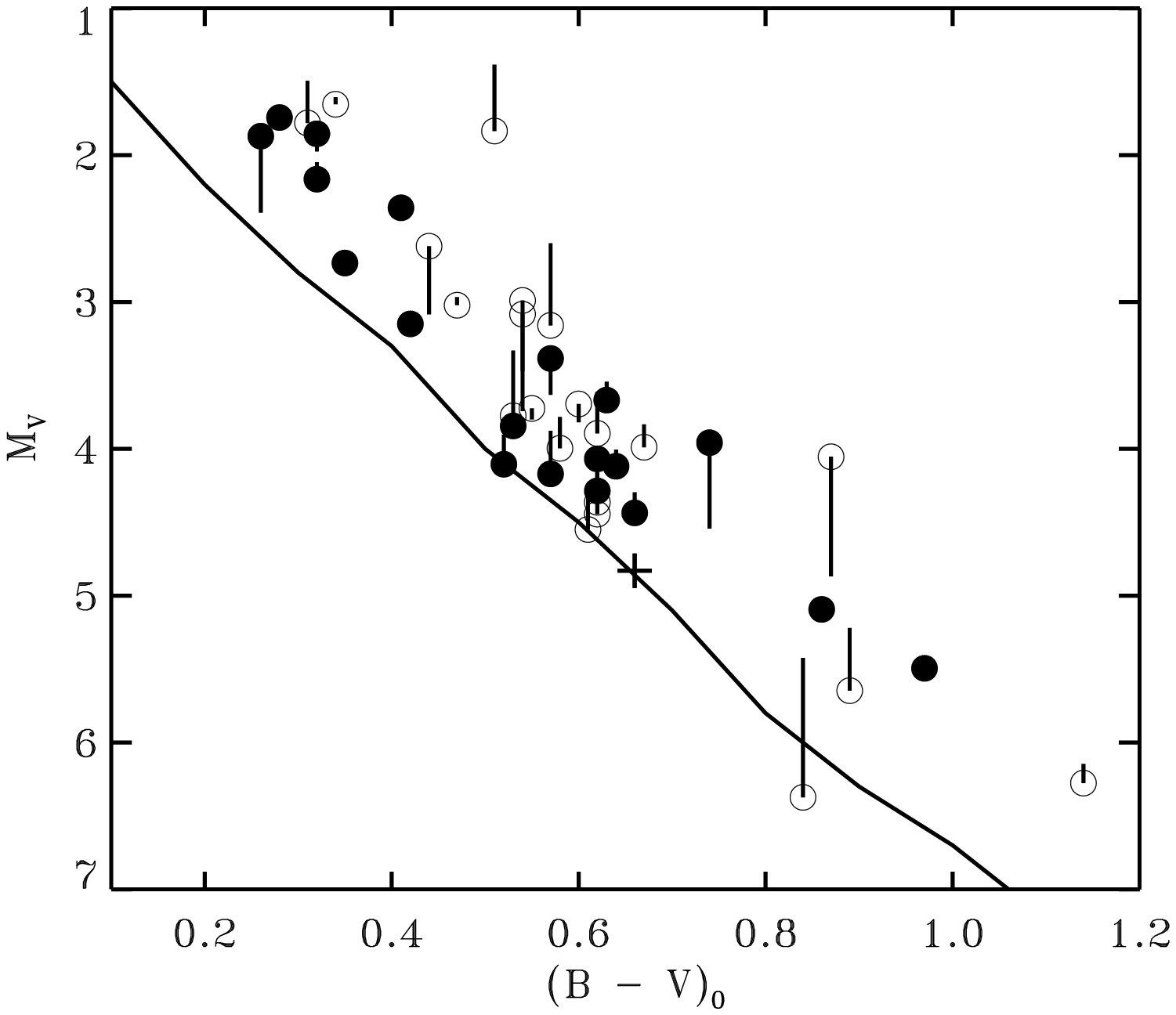,height=4.5in}}
\vskip 0.5in
\caption{\label{fig1} The color -- magnitude 
relation for the Hipparcos systems
with absolute-magnitude errors smaller than 0.25 magnitude (filled
circles) and within 0.25 to 0.5 magnitude (open circles). The same
symbols are used in all subsequent figures.
The line gives the main sequence for
single solar-type stars, with the Sun marked by a cross. The short
vertical vectors point at locations predicted by the new calibration
presented in this paper called CAL5.}
\end{figure}

The color -- absolute-magnitude diagram for the systems used in this paper
is shown in Figure~\ref{fig1}. Note the vertical spread of the points,
which is mostly due to the 
evolution of some systems away from the main sequence and to 
un-accounted mass-ratio differences. Without this spread, 
the calibration could be
simplified to just one independent variable. The period -- color
relation is shown in Figure~\ref{fig2}. In these and
the following figures, filled
circles mark the systems with better parallax data ($\epsilon M_V <
0.25$) whereas open circles mark the systems with larger relative
errors in the parallaxes ($0.25 < \epsilon M_V < 0.5$). There are
almost exactly the same numbers of systems in each of these
categories.

\placefigure{fig1}

\placefigure{fig2}

The combined 3-dimensional period -- color -- luminosity relation is
shown in Figure~\ref{fig3}. The planar fit corresponds to the 
calibration
\begin{eqnarray}
M_V & = & a_{P(BV)}\,\log P + a_{BV}\,(B-V)_0 + a_{0(BV)},   \label{eq5}
\end{eqnarray}
whose coefficients are listed in Table~\ref{tab2}. The
deviations from the fit are shown in Figure~\ref{fig4}.
As in the previous cases, the coefficients in Eq.~\ref{eq5} have errors
which are very large and very strongly non-Gaussian. In CAL1 and in
the following studies, the errors were estimated with a technique
of ``bootstrap re-sampling'', which utilizes a large number of solutions
based on the observational data re-arranged randomly
into several data sets of the same size by a random (and 
repetitious) selection of points. This technique permits to 
evaluate realistic errors for cases of
their entirely unknown distributions. The application to our data
is illustrated in Figure~\ref{fig5} which gives the spread
in the coefficients obtained from 10,000 bootstrap re-sampling solutions.
The one-sigma ranges 
in the diagrams can be identified with the extent of contours
containing 68.3\% of all cases. The essential statistics for the
bootstrap re-sampling results is given in Table~\ref{tab2}. Note 
that the median and average values of the coefficients are
closer than in the previous calibrations, indicating a better overall
stability of the current solution. Since this calibration is based on
a larger sample than all previous ones, Figure~\ref{fig5} no longer
shows multiple ``islands'' of the solutions which indicated
that the previous versions of the calibration were sensitive to the 
inclusion or rejection of individual systems.

\placetable{tab2}

\placefigure{fig3}

\placefigure{fig4}

\placefigure{fig5}

The relatively large uncertainties in the coefficients do not signify
that the predicted values of absolute
magnitudes are equally uncertain, as inter-parametric correlations
result in cancelation of the contributing uncertainties. 
To determine uncertainties in the predicted absolute magnitudes, a
Monte Carlo experiment was carried out. This experiment is based
on the availability of the 10,000 bootstrap solutions which were
used for evaluation of the coefficient errors, as described above.
We added one more level of randomization here by considering
100 random values of $(B-V)$ 
for each value of the period between 0.2 and 1.0 days in steps of 0.01
day. This way, {\it for each step in the period, $10^6$ random 
combinations of the colors and coefficients\/} were used.
Whereas the distribution of the coefficients had been given by the 
bootstrap re-sampling experiments, the 
distribution of the colors was assumed flat within period-dependent
ranges. These ranges increased linearly in width
from $\Delta (B-V) = 0.3$ at the period of 0.2 day to $\Delta (B-V) = 0.6$
 at the period of one day. The widening was meant to represent
the observed trend in the period -- color relation, with a much
larger spread in colors for long-period system. This spread
is usually interpreted as due to evolutionary effects 
which are more pronounced for long-period
systems. The blue edges of the respective color ranges were placed
at the observed blue, short-period envelope (BSPE) of the period -- color
distribution. The existence and importance of the BSPE was
discussed in CAL3 where an approximate fitting formula was determined: 
$(V-I)_{BSP} = 0.053\,P^{-2.1}$. Here, we used the same
BSP envelope, but transformed to the $(B-V)$ color 
using the Bessell\markcite{bes79} (1979) relations. Parenthetically,
we note that a more careful scrutiny of the
photometric data than in CAL4 resulted in a better definition of the
period -- color relation. Except for the
short-period system RW~Com, all systems are located within the band
marked in Figure~\ref{fig2}.

\placefigure{fig6}

The expected spread in the predicted values of $M_V$ for CAL5,
estimated from the Monte Carlo experiment described above, 
is shown in Figure~\ref{fig6}. The continuous and
broken lines give the spread in absolute magnitudes
encompassing 68.3\% and 95.4\% of all cases; these intervals
can be identified with one- and two-sigma ranges. As described
above, the spread stems from the combined uncertainties in the
coefficients and from the observed spread in $(B-V)$. It is
surprisingly small indicating that, in principle, the standard mean
error of predictions of $M_V$ should be at the level of 
0.1. The observed mean error is larger, $\sigma = 0.22$
(vertical bar in Figure~\ref{fig6}), indicating that the Monte Carlo
estimated errors may give overly optimistic estimates. Of
particular importance are three deviating systems with good
parallaxes, V759~Cen, SW~Lac and TY~Men, which are clearly too bright
than predicted by the calibration and which mainly contribute to the
increase of the mean error $\sigma$. 
 We have checked the input data for
these three systems and see no obvious reason why they deviate and
what is causing the high luminosities, since they also do not give
any clear hints of a third, hidden 
parameter dependence, as will be discussed in Section~\ref{third}.

\section{Do we see a third parameter in the luminosity calibration?}
\label{third}

The new luminosity -- period -- color relation may be accurate enough
to reveal systematic deviations of calibrating systems 
with properties deviating from the general trend. It was stated in CAL1
that the mass-ratio $q$ would be the most likely  
third parameter in the calibration. 
Thus, we may expect that the calibration should be fully 
valid for systems with an ``average
value'' of $q$, and that systems with mass-ratios deviating
noticeably from the average value would
show systematic residuals $\Delta
M_V$ from the calibrating plane. In addition to $q$, the fill-out
factor $f$ and the orbital inclination $i$ may 
influence the luminosity to a certain extent. The first two, $q$ and
$f$, are related to structural properties and are known
(Maceroni and van't Veer \markcite{mac96} 1996)
to correlate with the spectral type and total mass (early type systems 
have usually small mass ratios and large fill-out factors), so that 
we might expect that their contributions are 
already absorbed in the color term.

\placefigure{fig7}

We have collected the system parameters ($q$, $f$, $i$) for the 
majority of the systems used in CAL5. The data are given in
Table~\ref{tab1}, details on the source of data are given in the
Appendix. The residuals  $\Delta M_V$ from the calibration are plotted 
versus each of these parameters in Figure~\ref{fig7}. The formal
linear fits, giving coefficients as in
Table~\ref{tab3}, indicate no significant dependencies for $q$ and
$f$; this table gives also results of bootstrap re-sampling
experiments. The lack of  
systematic trends is also obvious in Figure~\ref{fig7}.
Only the $i$ dependence shows a 0.15 mag variation
in the interval $60^\circ$ to $90^\circ$ which is well covered by
observed systems, indicating the possibility of
a gravity-induced brightening at the poles.
The slope per degree, $+0.0053^{+0.0027}_{-0.0041}$, is significant at
the level slightly higher than one-sigma.
Since this effect is of the order of the accuracy of the calibration,
it can be neglected in statistical studies.

\placetable{tab3}

In addition to the geometrical parameters ($q$, $f$, $i$) one could
consider metallicity as a hidden unknown of the calibration. Unfortunately,
nothing is known about differences in metallicities within the
sample of the local W~UMa-type systems. Similarly, 
usage of any correlated indicators, 
such as spatial velocities, would not be advisable
as one would expect a statistical rather 
than a deterministic relation between deviations in $M_V$ and the
values of the space velocity vectors.

\section{The poor thermal contact and semi-detached systems}
\label{poor}

As was described in CAL1 and CAL4, the existence of 
a period -- color relation causes an insufficient separation of the 
color and period dependencies so that the results crucially
depend on the data for ``outlying'' systems which
are contact systems with unusually long
periods for their colors. Such systems appear close to the
red-color (broken line) boundary in Figure~\ref{fig2}.
 They are important as they can stabilize
the solutions by lifting the period -- color degeneracy. For the
previous calibrations in CAL1 to CAL4, the system V371~Cep (NGC~188
V5) was used in this special role. The current Hipparcos 
sample defined by $\epsilon M_V < 0.5$ includes a wide range of
combinations of periods and colors, so that it is no longer necessary 
to rely on extreme systems such as V371~Cep. In fact, we are for
the first time in a position to check if the calibration is indeed valid
for such systems. As indicated by the OGLE data (Rucinski 
\markcite{97b}1997b), the poor-thermal-contact (PTC) and possibly 
semi-detached (sd) systems  mimicking good contact
are very rare in space, constituting only some 2\% of all contact systems.
They are very important for our understanding of pre- and
in-contact evolution. We used V371~Cep in the previous versions of the
luminosity calibration to improve stability of solutions,
but {\it it has never been proven that the
luminosity calibrations of the CAL series are 
applicable to such systems at all!\/}. 

There are very few binaries with reliable values of $M_V$ among the
PTC/sd systems. In addition to V371~Cep, one can consider here RT~Scl
which seems to be a genuine semi-detached system (Hilditch and King
\markcite {hil86} 1986, Banks et al.\ \markcite{ban90} 1990), perhaps
also the red system RW~Com 
(Milone et al.\ \markcite{mil87} 1987) which is included
in our calibration, although with very low weight.  
If we apply the new calibration to such systems,
their absolute visual magnitudes (derived from
spectroscopic and photometric analysis) are usually fainter than
predicted. The deviations amount to 0.80 magnitude for V371~Cep,
0.40 magnitude for RT~Scl, while RW~Com deviates by as much as 0.95
magnitude. 

The sample of the PTC/sd systems is clearly too small to draw definite
conclusions, but we point out that low luminosities (for a given
color) are actually expected for such systems. By comparing mixing of
colors of Main Sequence stars (clearly the extreme case of no thermal
contact) with colors of genuine contact systems, one sees that the
PTC/sd systems should be always bluer than systems in good geometrical
and thermal contact (Figure~\ref{fig8}). The observed colors will
obviously depend on how good or poor the thermal contact actually is, so
that their values cannot be predicted easily. But the result is clear:
The luminosities estimated from the calibration are expected to
be always too high, which seems to be confirmed by results described
above. It would be of great importance to confirm this prediction on the
basis of a larger sample of PTC/sd systems.

\placefigure{fig8}

\section{Conclusions}
\label{concl}
We have presented a luminosity calibration of contact binaries based on
Hipparcos parallaxes, which covers a wide range in periods and colors.
It is based on the $(B-V)$ color index and supersedes previous
calibrations based on that color index. It nominally predicts 
absolute magnitudes to an accuracy of $\pm 0.1$, although deviations
of some systems are larger, indicating a wide, non-Gaussian
distribution of uncertainties. For the current data, the weighted mean
error is $\sigma = 0.22$. The calibration can be written as:
\setcounter{equation}{4}   
\alpheqn
\begin{equation}
M_V = -4.44 \log P + 3.02 (B-V)_0 +0.12
\end{equation}
We have investigated the influence of mass-ratio, fill-out factor and
inclination, and have found that the first two parameters are,
within the accuracy of the calibration, completely absorbed by the
color term. Only the inclination, which is independent of the 
physical configuration of a system, shows a noticeable effect, as
it is expected from the gravity brightening. Nevertheless, this
effect is so small that it can be neglected in statistical studies.
Close binary systems appearing in the period domain of the W~UMa-type
systems, but showing unequally deep eclipses (either
poor-thermal contact or semi-detached systems) appear too faint
relative to the calibration by roughly 0.5 magnitude, but the sample
is much too small to draw definite conclusions.

The accuracy of the present calibration is no longer limited by the
parallax data but, paradoxically, by the lack of reliable photometric
data. It should be noted that an uncertainty in the color index
$(B-V)$ of 0.03 contributes a deviation of 0.1 magnitude in the
predicted $M_V$. 

HWD gratefully acknowledges support from the sabbatical 
visitors program at Space Telescope Science Institute 
and SMR acknowledges a research grant from the Natural 
Sciences and Engineering Council of Canada. The research has made use
of the SIMBAD database, operated at the CDS, Strasbourg, France.

\section*{Appendix: Sources of input data}

In the following, information is given on the source of apparent 
magnitudes at maximum, color indices and system parameters of the
W~UMa systems used in CAL5.

\noindent
{\bf AB~And.} $V$ magnitude and $(B-V)$ index from Landolt 
\markcite{lan69}(1969). He quotes  $(B-V)=0.90$ as the color index at 
total eclipse. Using the color index of the comparison, and differential
colors of the contact system, we get 0.865 for the (fairly poorly
determined) maximum and 0.90 for the minimum. The $(b-y)$-index 
suggests $(B-V)=0.88$, which was adopted for CAL5. The system parameters
were taken from the detailed study of Hrivnak \markcite{hri88}(1988), 

\noindent
{\bf S~Ant.} $V$ and $(B-V)$ from Popper \markcite{pop57}(1957),
system parameters from Russo et al.\ \markcite{rus82}(1982).
 
\noindent
{\bf V535~Ara.} $V$ and $(B-V)$ from Sch\"offel
\markcite{sch79}(1979), system parameters 
are averages of the values given by Sch\"offel \markcite{sch79}(1979)
and Leung and \markcite{leu78}Schneider (1978).
 
\noindent
{\bf TZ~Boo}. This system shows a very unstable light curve. $V$ and
$(B-V)$ averaged from data by Hoffmann \markcite{hof78}(1978), mass 
ratio given by McLean and Hilditch \markcite{mcl83}(1983), inclination
estimated by Al-Naimiy and Jabbar \markcite{aln87}(1987).
 
\noindent
{\bf AC~Boo.} $(B-V)$ given by Mauder \markcite{mau64}(1964). Several 
$B,\,V$ light curves are available, but none refers to a comparison star
with known $B,\,V$ magnitudes. The $V$ magnitude at maximum was 
calculated from the $Hp$ magnitude. Mass ratio from Hrivnak 
\markcite{hri93}(1993), inclination and fill-out factor from Mancuso
et al.\ \markcite{man78}(1978).
 
\noindent
{\bf CK~Boo.} Aslan and Derman \markcite{aslasl86}(1986) measured 
{\it UBV} light curves relative to HD~128128. Str\"omgren photometry
of HD~128128 (Olsen \markcite{ols94}1994) yields $V=7.897$,
$b-y=0.348$, from which $(B-V)=0.55$ was derived. Mass ratio 
from Hrivnak \markcite{hri93}(1993).
 
\noindent
{\bf RR~Cen.} $V$ and $(B-V)$ from Chambliss \markcite{cha71}(1971), 
system parameters averaged from Mochnacki and Doughty
\markcite{moc72}(1972) and King and Hilditch \markcite{kin84}(1984). 
 
\noindent
{\bf V752~Cen.} $V$ and $(B-V)$  from Sistero and Castore de 
Sistero \markcite{sis73}(1973), also $(B-V)$ from Duerbeck
\markcite{due97}(1997). Averaged color index used here. System 
parameters taken from Barone et al.\ \markcite{bar93}(1993).
 
\noindent
{\bf V757~Cen.} $V$ and $(B-V)$ from Cerruti and Sistero
\markcite{cer82}(1982), also $(B-V)$ from Duerbeck
\markcite{due97}(1997). Averaged color index used here.  
System parameters taken from Maceroni, Milano and Russo
\markcite{mil84}(1984). 
 
\noindent
{\bf V759~Cen.} $(B-V)$ was derived from Bond's \markcite{bon70}(1970)
$b-y$ value. Maximum $V$ magnitude was calculated from the $Hp$ magnitude.
 
\noindent
{\bf VW~Cep.} Linnell \markcite{lin82}(1982) gives $B,V$ light curves 
relative to BD~$+74^\circ\ 889$, whose $B,\,V$ magnitudes were taken
from Sorgsepp and Albo \markcite{sor74}(1974). System parameters were
taken from Hill \markcite{hil89}(1989) and Mochnacki \markcite{moc81}(1981).
 
\noindent
{\bf RS~Col.} $V$ and $(B-V)$ from McFarlane and Hilditch 
\markcite{mcf87}(1987), who also determined system parameters.
 
\noindent
{\bf RW~Com.} $V$ and $(B-V)$ from Milone et al.\ \markcite{mil80}(1980), 
system parameters from Milone, Wilson and Hrivnak \markcite{mil87}(1987).
 
\noindent
{\bf $\epsilon$~CrA.} $V$ and $(B-V)$ from Tapia \markcite{tap69}(1969), 
mass ratio from Goecking and Duerbeck \markcite{goe93}(1993), inclination 
and fill-out factor by Tapia and Whelan \markcite{tap}(1975).
 
\noindent
{\bf SX~Crv.} $V$ and $(B-V)$ are averaged from Sanwal et al.\
\markcite{san74} (1974) and Scaltriti and Busso\markcite{sca84} (1984). 
 
\noindent
{\bf V1073~Cyg.} The $(B-V)$ index is given by Mendoza, G\'omez, \& 
Gonz\'alez \markcite{men78}(1978). Several light curves
are available, but none refers to a comparison star with known $B,\,V$ 
magnitudes. The $V$ magnitude at maximum was calculated from the $Hp$ 
magnitude. The system parameters were taken from the study of Ahn, Hill and
Khalesseh \markcite{ahn92}(1992).
 
\noindent
{\bf RW~Dor.} $V$ and $(B-V)$ from Marton, Grieco and Sistero
\markcite{mar89}(1989), who also give system parameters. The
parameters used here are average values of the previous study and that
of Kaluzny and Caillault \markcite{kal89}(1989). 
 
\noindent
{\bf AP~Dor.} $V$ magnitude at maximum estimated from Eggen's
\markcite{egg80}(1980) observations, $(B-V)$ taken from Przybylski and
Kennedy \markcite{65}(1965). 
 
\noindent
{\bf BV~Dra} and {\bf BW Dra}. $V$  and $(B-V)$ averaged from Yamasaki
\markcite{yam79}(1979) and Rucinski and Kaluzny
\markcite{ruc82}(1982). The system parameters were taken from
Kaluzny and Rucinski \markcite{kal86}(1986). 
 
\noindent
{\bf YY~Eri.}  $V$ at maximum calculated from the $Hp$ magnitude.
$(B-V)$ taken from Eggen \markcite{egg67}(1967). System parameters
were derived by Nesci et al.\ \markcite{nes86}(1986).
 
\noindent
{\bf SW~Lac.} $V$ and $(B-V)$ from Rucinski \markcite{ruc68}(1968). 
System parameters were derived by Leung, Zhai and Zhang \markcite{leu84}(1984).
 
\noindent
{\bf XY~Leo.} $V$ from Hrivnak \markcite{hri85}(1985). The $(B-V)$
index is an average of the determinations of Koch
\markcite{koc74}(1974), Hrivnak \markcite{hri85}(1985), and Duerbeck
\markcite{due97}(1997). System parameters were derived by Hrivnak
\markcite{hri85}(1985) and Barden \markcite{bar87}(1987). 
The latter found that XY~Leo is a quadruple system. The $V$ magnitude
has been increased by 0.05 to account for the M-type pair.
 
\noindent
{\bf AP~Leo.} $V$ at maximum calculated from the $Hp$ magnitude.
$(B-V)$ taken from Duerbeck \markcite{due97}(1997). System parameters
were derived by Zhang, Zhang and Zhai \markcite{zha92}(1992).
 
\noindent
{\bf UV~Lyn.} $V$ and $(B-V)$ from Bossen (1973). A more recent 
derivation of system parameters was carried out by Markworth and Michaels 
\markcite{mar82}(1982).
 
\noindent
{\bf TY~Men.} Naqvi and Gr\o{}nbech \markcite{naq76}(1976) obtained 
{\it uvby} light curves. $V$ magnitude averaged from both maxima; 
$(B-V)$ calculated from average $(b-y)$ index. The system parameters 
were derived by Lapasset \markcite{lap80}(1980).
 
\noindent
{\bf UZ~Oct.} $V$ and $(B-V)$ taken from Castore de Sistero, Sistero and 
Candellero \markcite{cas79}(1979).  The system parameters were derived
by Lapasset and Sistero \markcite{lap84}(1984).
 
\noindent
{\bf V502~Oph.} $V$ and $(B-V)$  taken from Wilson
\markcite{wil67}(1967). System parameters are averaged from results 
of King and Hilditch \markcite{kin84}(1984) and Maceroni et 
al.\ \markcite{mac82}(1982).
 
\noindent
{\bf V566~Oph.} Lafta and Grainger \markcite{laf85}(1985) give $B,\,V$
photometry relative to BD $+4^\circ\ 3558$, whose {\it UBVRI} photometry 
is given by Castelaz et al.\ \markcite{cas91}(1991). System parameters
are averaged from results of Van Hamme and Wilson \markcite{Van85}(1985) 
and Mochnacki and Doughty \markcite{moc72}(1972).
 
\noindent
{\bf V839~Oph.} Several $B,V$ light curves are available, but none refers 
to a comparison star with known $B,\,V$ magnitudes. $V$ at maximum was 
derived from the Hp magnitude. $(B-V)$ is an average of the value
published by Koch (1974) and that calculated from the $(b-y)$ index
determined by Rucinski and Kaluzny \markcite{ruc81}(1981). The 
inclination is taken from an analysis by Niarchos \markcite{nia89}(1989).
 
\noindent
{\bf MW~Pav.} $V$ and $(B-V)$ from Lapasset \markcite{lap77}(1977), 
system parameters taken from Lapasset \markcite{lap80}(1980).
 
\noindent
{\bf U~Peg.} No light curve relative to a comparison star with known
$B,\,V$ magnitudes exists. $V$ at maximum was derived from the 
$Hp$ magnitude. $(B-V)$ was taken from Eggen \markcite{egg67}(1967), 
system parameters from the study of Zhai and Lu \markcite{zha88}(1988).
 
\noindent
{\bf AE~Phe.} $V$ and $(B-V)$ from Walter and Duerbeck
\markcite{wal88}(1988), system parameters were averaged from the 
results of Van Hamme and Wilson \markcite{Van85}(1985) and Niarchos
and Duerbeck \markcite{nia91}(1991). 
 
\noindent
{\bf VZ~Psc.} $V$ at maximum was extrapolated from Bradstreet's 
\markcite{bra85}(1985) fragmentary {\it UBVRI} observations. 
Several measurements of $(B-V)$ are available; a ``median'' value is 
that measured by Ryan \markcite{rya89}(1989).
System parameters were determined by Hrivnak, Guinan and Lu
\markcite{hri95}(1995). 
 
\noindent
{\bf AQ~Psc.} $V$  and $(B-V)$ from Sarma and Radhakrishnan
\markcite{sar82}(1982). 
 
\noindent
{\bf HI~Pup.} $V$ at maximum was calculated from the $Hp$ magnitude,
$(B-V)$ was taken from Duerbeck \markcite{due97}(1997).
 
\noindent
{\bf V781~Tau.} $V$ and $(B-V)$ from Cereda et al.\ \markcite{cer88}(1988). 
System parameters were determined by Lu \markcite{lu93}(1993).
 
\noindent
{\bf W~UMa.} We used the $B,\,V$ light curves by Breinhorst
\markcite{bre71}(1971), which were measured relative to BD~$+56^\circ\ 1399$,
for which Oja \markcite{oja84}(1984) gives {\it UBV} magnitudes. 
System parameters were taken from the study of Linnell \markcite{lin91}(1991).
 
\noindent
{\bf AW~UMa.} $V$ and $(B-V)$ from Paczynski \markcite{pac64}(1964). 
System parameters were averaged from the studies by Rucinski 
\markcite{ruc92}(1992) and Hrivnak \markcite{hri82}(1982).
 
\noindent
{\bf GR~Vir.} $V$ and $(B-V)$ from Cereda et al.\ \markcite{cer88}(1988).

\begin{figure}           
\centerline{\psfig{figure=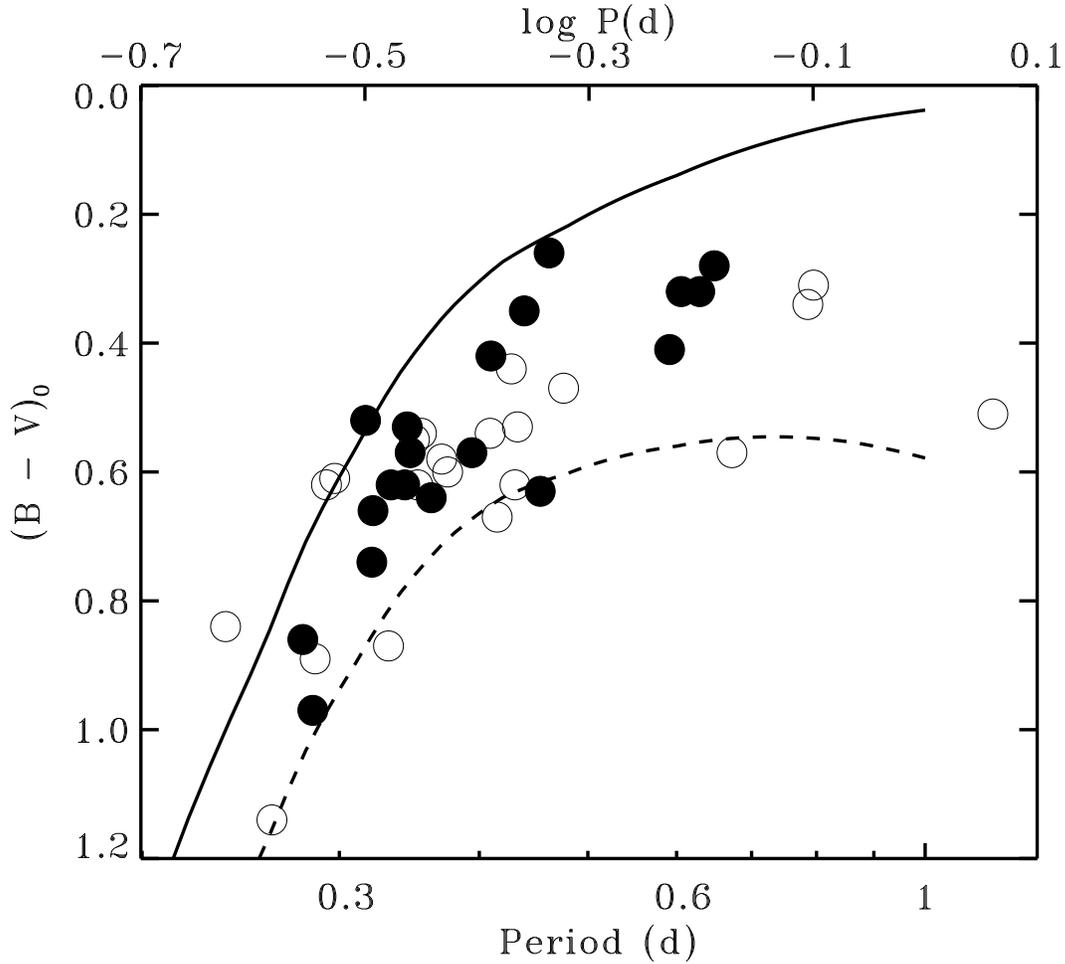,height=4.5in}}
\vskip 0.5in 
\caption{\label{fig2}
The period -- color diagram for the W~UMa-type systems observed by
Hipparcos (filled and open circles, as in Figure~1). The solid line is the
blue--short-period envelope (BSPE) from CAL3, transformed to the
$(B-V)$ color. The broken line gives the red edge of the band 
which was used in a Monte Carlo experiment to estimate 
the spread in $M_V$. Its location was roughly estimated
to include evolved systems with red colors
(for a given period). The red edge was constructed by adding a 
linearly-increasing value to the BSPE, so that the band 
would widen from 0.3 at 0.2 days to 0.6 at 1.0 day.
The BSPE is poorly defined for periods longer than about 0.6 -- 0.7 day.}
\end{figure}

\begin{figure}           
\centerline{\psfig{figure=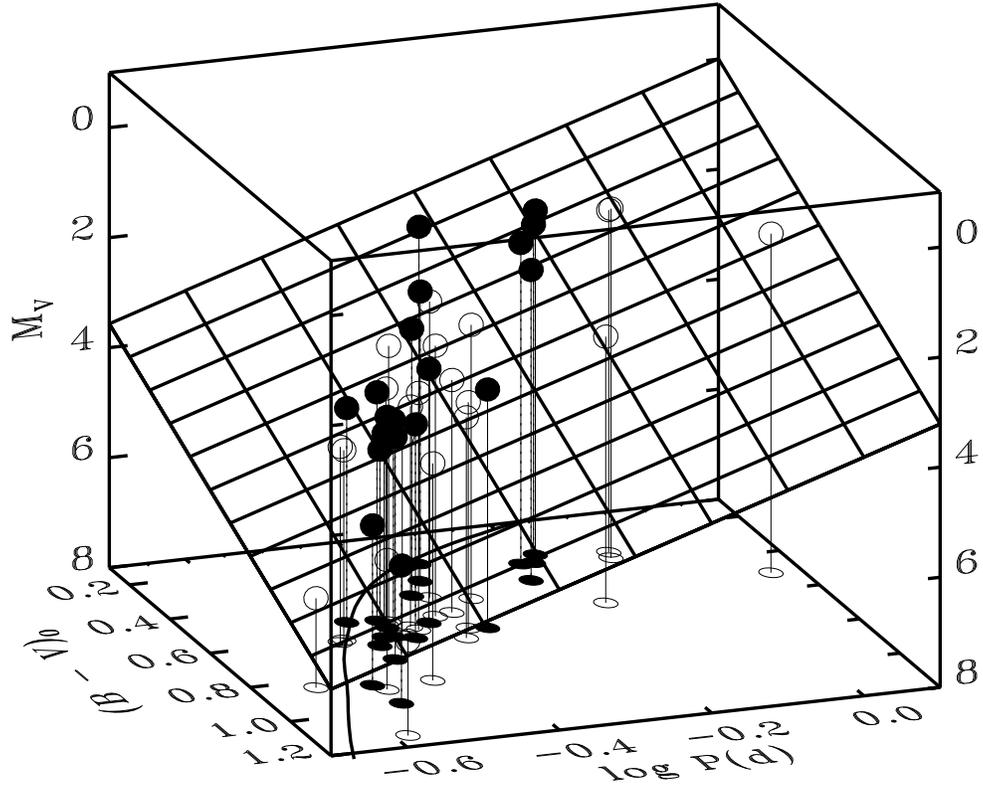,height=4.5in}}
\vskip 0.5in 
\caption{\label{fig3}
The period -- color -- luminosity relation for the Hipparcos
systems. The inclined surface gives the new calibration CAL5. The mean
weighted deviation of the observed points from the calibration plane
is $\sigma = 0.22$. The data projected onto the horizontal 
period -- color plane yield the same period -- color 
relation as in Figure~2.}
\end{figure}

\begin{figure}           
\centerline{\psfig{figure=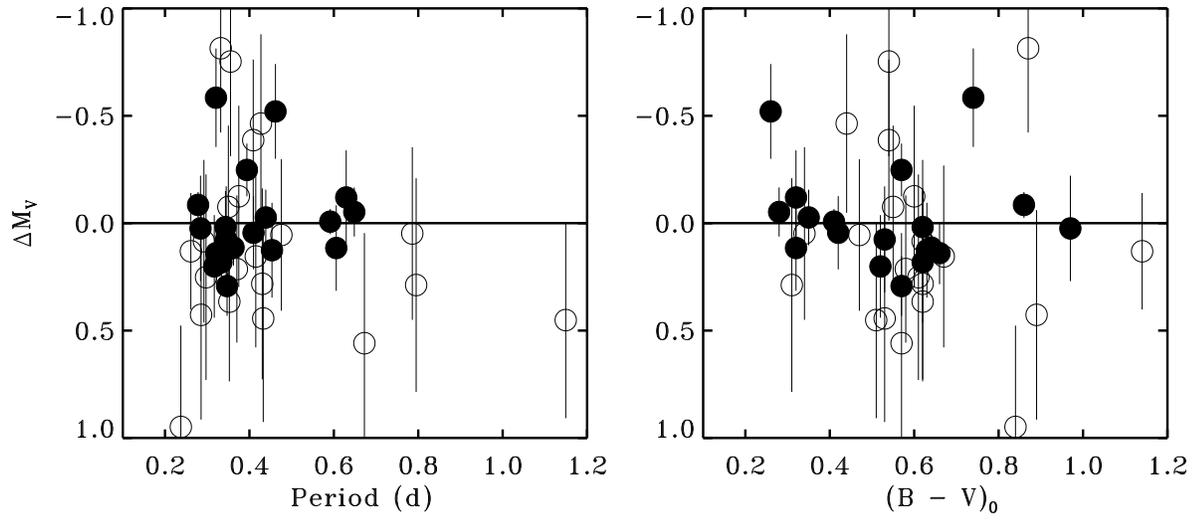,height=4.5in,angle=90}}
\vskip 0.5in 
\caption{\label{fig4}
Deviations from the new calibration CAL5 are shown here as functions
of the two principal parameters. The vertical error bars have lengths
equal to twice the absolute-magnitude errors, as estimated from the
uncertainties in the Hipparcos parallaxes.}
\end{figure}

\begin{figure}           
\centerline{\psfig{figure=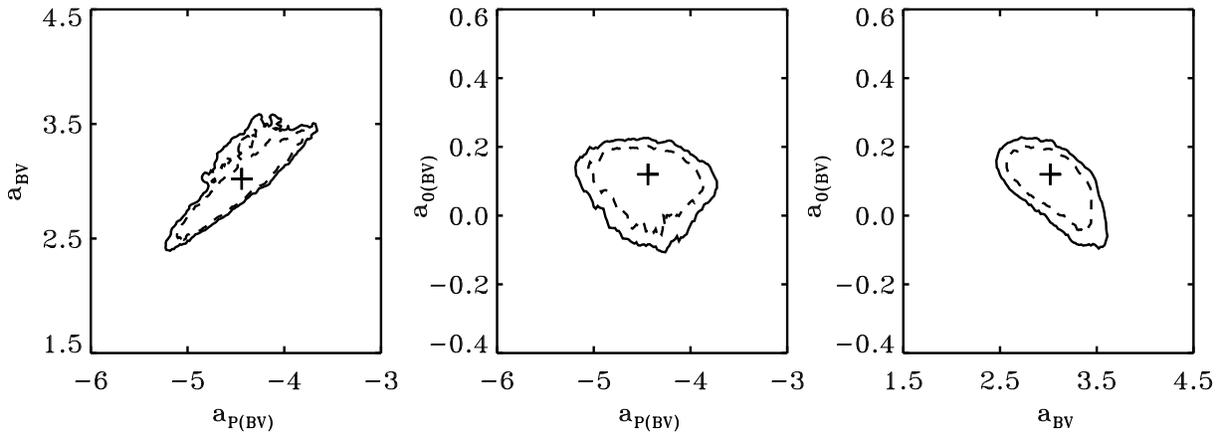,height=4.5in,angle=90}}
\vskip 0.5in 
\caption{\label{fig5}
Relations between the coefficients of the new calibration CAL5
established by a bootstrap re-sampling experiment. 
The naming of the coefficients is the same as
in CAL1, in that BV in parentheses indicates that this is a $(B-V)$-based
calibration. 
The solid contours encompass 68.3\% of all bootstrap solutions of the
coefficients, a level which is normally associated with the one-sigma
standard-error uncertainty. The dotted lines give similar levels 
for 50\%, which is normally associated with the probable errors. Note
that separate ``islands'' in mutual correlations between the
coefficients -- which were present in the previous calibrations --
no longer exist indicating higher stability of the present solutions.}
\end{figure}

\begin{figure}           
\centerline{\psfig{figure=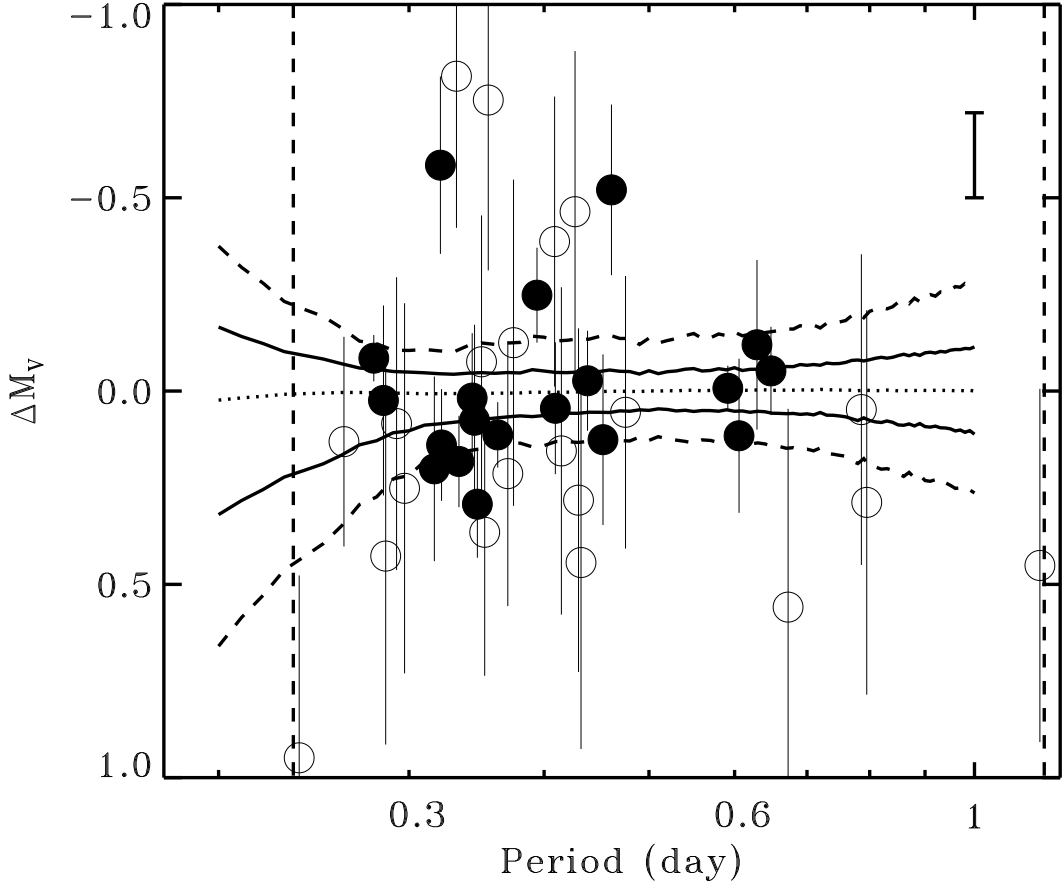,height=4.5in}}
\vskip 0.5in 
\caption{\label{fig6}
The spread in the predicted values of $M_V$ obtained from  
a Monte Carlo experiment for CAL5 is shown here by curves. It was derived
by combining the spread in the calibration
coefficients, as shown in Figure~5, with an assumed spread in colors.
In practice, 10,000 calibration coefficients obtained in 
the bootstrap experiments were applied to
100 uniformly  distributed random 
values of the colors (within ranges shown in Figure~2)
for each value of the period in steps of 0.01 day.
The solid lines give the one-sigma range (68.3\%
of all cases), whereas the broken lines give the two-sigma range
(95.4\% of all cases). The median for the random samplings is shown by
the dotted line. The two vertical broken 
lines delineate the region defined by systems with the extreme 
values of the orbital period in the Hipparcos sample. The vertical bar
gives the weighted mean deviation in the sample, $\sigma = 0.22$. 
The filled and open circles 
as well as the vertical bars have the same meanings as 
in other figures.}
\end{figure}

\begin{figure}           
\centerline{\psfig{figure=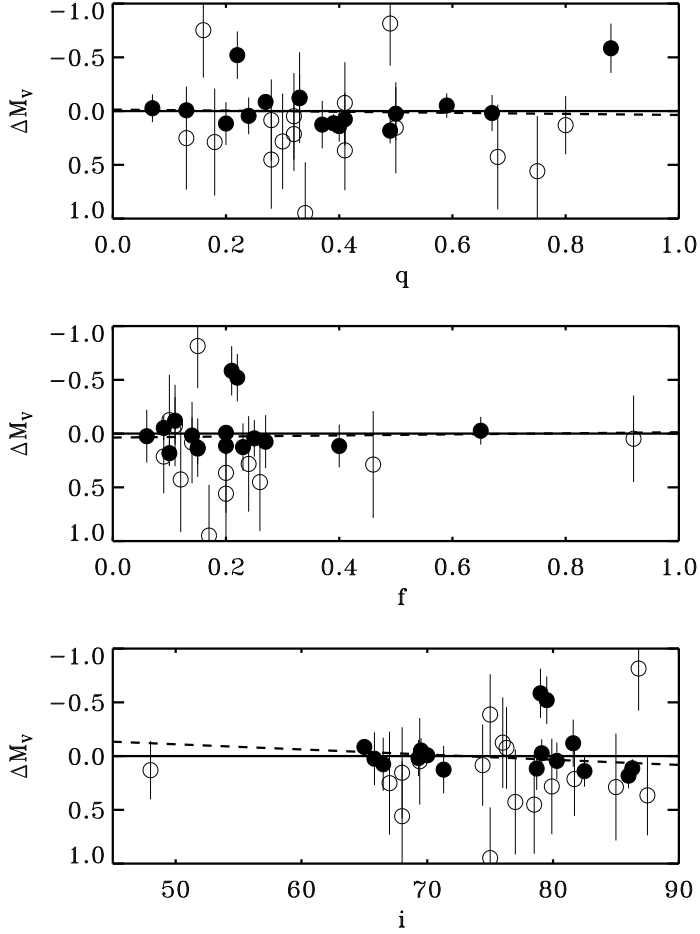,height=4.5in}}
\vskip 0.5in 
\caption{\label{fig7}
Deviations from the calibration are shown here versus the physical
parameters of the contact system: the mass ratio, $q = M_2/M_1 \le 1$,
the fill-out parameter, $f$, defined through 
equipotentials with $f=0$ for the inner common equipotential, and 
the orbital inclination, $i$, in degrees.
The filled and open circles as well as the vertical bars
have the same meanings as in previous figures. 
The dashed lines indicate the weighted least squares fits to the data.
Note that ($q$, $f$, $i$) 
data are not available for all systems studied here.}
\end{figure}

\begin{figure}           
\centerline{\psfig{figure=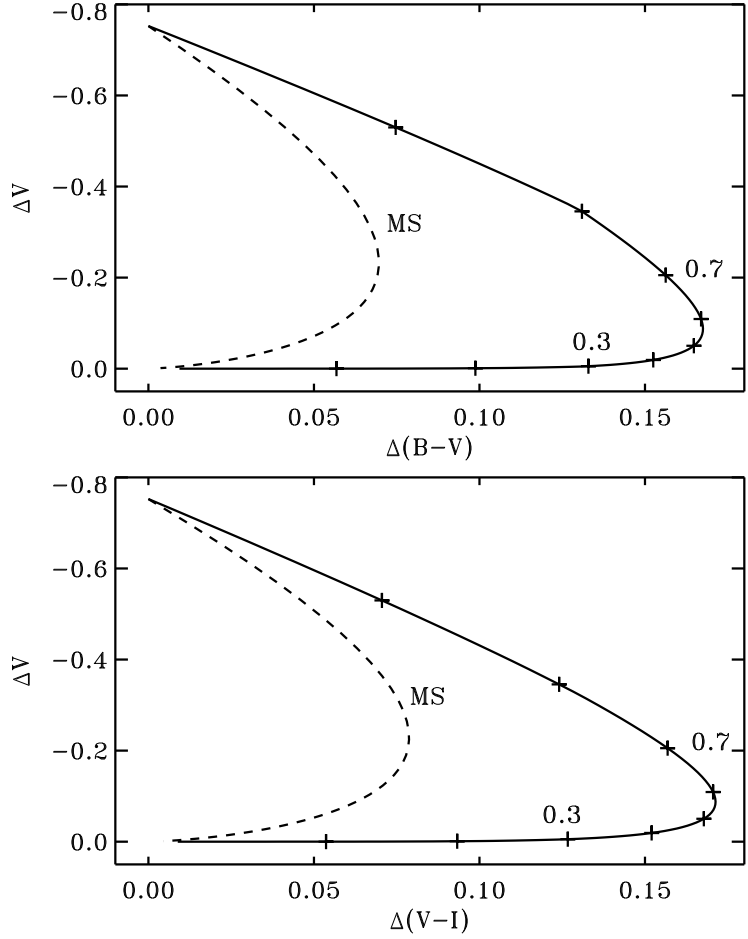,height=4.5in}}
\vskip 0.5in 
\caption{\label{fig8}
Expected changes in $V$ and $(B-V)$ (upper panel) and $V$ and $(V-I)$
(lower panel) for good-thermal contact systems
with the varying mass-ratio, $q$, are shown by the continuous lines
with marks every 0.1 in $q$. The broken lines show simple mixing of
colors for Main Sequence stars which can be taken as an 
extreme case of poor thermal contact. The calculations presented here
are for the color of the primary component $(B-V) =0.75$ and 
$(V-I) = 0.8$ and assume the slope of the Main Sequence $M_V \propto 4.85
(B-V)$ and $M_V \propto 4.35 (V-I)$. The color changes for good
thermal-contact systems have been
computed following the analysis of Mochnacki (1981) assuming the
nuclear mass
-- luminosity relation $L \propto M^{+4.4}$. Note that the largest
color changes are observed for $0.5 <
q < 0.6$, and that they are accompanied by very small changes in
brightness. Apparently, for such mass-ratios, the contact systems
have secondaries contributing large radiating areas, but almost no
luminosity. For $q \rightarrow 0$ the systems resemble single stars,
whereas for $q \rightarrow 1$ the stars again influence each other to
a lesser degree as they become identical, with the combined $\Delta V
\rightarrow  -0.75$. The poor thermal contact systems are expected to
be located between the continuous and broken curves in the figure.}
\end{figure}


\begin{deluxetable}{lrcccccccccccc}
\small
\tablenum{1}
\tablewidth{0pt}
\tablecaption{\label{tab1}
Contact binary stars observed by Hipparcos with
$\epsilon M_V \le 0.5$}
\tablehead{
\colhead{Name}       & \colhead{$P$}        &
\colhead{$B-V$}      & \colhead{$E_{B-V}$}  &
\colhead{$V_{\rm max}$}  & \colhead{$Hp_{\rm max}$} & 
\colhead{$q$}        & \colhead{$f$}        & 
\colhead{$i$}        & \colhead{$\pi$}      & 
\colhead{$\epsilon \pi$} & \colhead{$\epsilon M_V$} & 
\colhead{$M_V$}      & \colhead{$\Delta M_V$} }
\startdata
AB  And &  0.3319 &  0.88 &  0.01 &  9.49 &  9.68 &  0.49 &  0.15 &
86.6 &  \phn 8.3 &  1.5 &  0.39 & 4.05 & $-0.81$ \nl
S   Ant &  0.6484 &  0.33 &  0.05 &  6.28 &  6.37 &  0.59 &  0.09 &  
69.5 & 13.3 &  0.7 &  0.11 & 1.74 & $-0.05$ \nl
V535 Ara &  0.6293 &  0.34 &  0.02 &  7.17 &  7.24 &  0.33 &  0.11 &
81.6 & \phn 8.9 &  0.9 &  0.22 & 1.85 & $-0.12$ \nl
TZ  Boo &  0.2972 &  0.64 &  0.03 & 10.48\phn & 10.57\phn &  0.13 &\nodata &
67.0 & \phn 6.8 &  1.5 &  0.48 & 4.55 &  $+0.25$ \nl
AC  Boo &  0.3524 &  0.62 &  0.00 &  9.96 & 10.09\phn &  0.41 &  0.20 &  
87.5 & \phn 7.6 &  1.3 &  0.37 & 4.36 &  $+0.37$ \nl
CK  Boo &  0.3552 &  0.55 &  0.01 &  8.99 &  9.09 &  0.16 &\nodata&
\nodata& \phn 6.4 &  1.3 &  0.44 & 2.99 & $-0.75$ \nl
RR  Cen &  0.6057 &  0.34 &  0.02 &  7.27 &  7.40 &  0.20 & 0.40 &  
78.7 & \phn 9.8 &  0.9 &  0.20 & 2.16 & $+0.12$ \nl
V752 Cen &  0.3702 &  0.60 &  0.02 &  9.17 &  9.20 &  0.32 &  0.09 &  
81.7 & \phn 9.5 &  1.5 &  0.34 & 4.00 & $+0.21$ \nl
V757 Cen &  0.3432 &  0.65 &  0.03 &  8.40 &  8.50 &  0.67 &  0.14 &  
69.3 & 14.2 &  1.1 &  0.17 & 4.07 & $+0.02$ \nl
V759 Cen &  0.3939 &  0.59 &  0.02 &  7.44 &  7.56 &\nodata&\nodata&
\nodata& 15.9 &  0.9 &  0.12 & 3.38 & $-0.25$ \nl
VW  Cep &  0.2783 &  0.86 &  0.00 &  7.30 &  7.54 &  0.27 & 0.00 &  
65.0 & 36.2 &  1.0 &  0.06 & 5.09 & $-0.09$ \nl
RS  Col &  0.6724 &  0.59 &  0.02 &  9.52 &  9.61 &  0.75 & 0.20 &  
68.0 & \phn 5.5 &  1.3 &  0.51 & 3.16 & $+0.56$ \nl
RW  Com &  0.2375 &  0.84 &  0.00 & 11.07\phn & 11.24\phn &  0.34 & 0.17 &  
75.0 & 11.5 &  2.5 &  0.47 & 6.37 & $+0.95$ \nl
$\epsilon$ CrA &  0.5914 &  0.41 &  0.00 &  4.74 &  4.82 &  0.13 &
0.20 &  70.0 & 33.4 &  0.9 &  0.06 & 2.36 & $-0.01$ \nl
SX  Crv &  0.3166 &  0.54 &  0.02 &  8.98 &  9.05 &\nodata&\nodata&
\nodata& 10.9 &  1.2 &  0.24 & 4.11 & $+0.20$ \nl
V1073 Cyg &  0.7859 &  0.42 &  0.08 &  8.24 &  8.33 &  0.32 &  0.92 &  
69.4 & \phn 5.4 &  1.0 &  0.40 & 1.65 & $+0.05$ \nl
RW  Dor &  0.2855 &  0.89 &  0.00 & 10.90\phn & 11.07\phn &  0.68 & 0.12 &  
77.0 & \phn 8.9 &  2.0 &  0.49 & 5.65 & $+0.43$ \nl
AP  Dor &  0.4272 &  0.44 &  0.00 &  9.26 &  9.37 &\nodata&\nodata&
\nodata& \phn 4.7 &  0.9 &  0.42 & 2.62 & $-0.46$ \nl
BV  Dra &  0.3501 &  0.56 &  0.01 &  7.89 &  7.94 &  0.41 &  0.11 &  
76.3 & 14.9 &  2.6 &  0.38 & 3.72 & $-0.08$ \nl
BW  Dra &  0.2922 &  0.63 &  0.01 &  8.61 &\nodata&  0.28 &  0.14 &  
74.4 & 14.9 &  2.6 &  0.38 & 4.44 & $+0.08$ \nl
YY  Eri &  0.3215 &  0.66 &  0.00 &  8.16 &  8.30 &  0.40 &  0.15 &  
82.5 & 18.0 &  1.2 &  0.14 & 4.44 & $+0.14$ \nl
SW  Lac &  0.3207 &  0.75 &  0.01 &  8.54 &  8.80 &  0.88 &  0.21 &  
79.0 & 12.3 &  1.3 &  0.23 & 3.96 & $-0.58$ \nl
XY  Leo &  0.2841 &  0.99 &  0.02 &  9.55 &  9.64 &  0.50 &  0.06 &  
65.8 & 15.9 &  1.8 &  0.25 & 5.49 & $+0.02$ \nl
AP  Leo &  0.4304 &  0.62 &  0.00 &  9.30 &  9.43 &  0.30 &  0.24 &  
79.9 & \phn 8.3 &  1.7 &  0.44 & 3.90 & $+0.28$ \nl
UV  Lyn &  0.4150 &  0.67 &  0.00 &  9.42 &  9.58 &  0.50 &\nodata &
68.0& \phn 8.2 &  1.6 &  0.42 & 3.99 & $+0.15$ \nl
TY  Men &  0.4617 &  0.29 &  0.03 &  8.11 &  8.17 &  0.22 &  0.22 &  
79.5 & \phn 5.9 &  0.6 &  0.22 & 1.87 & $-0.52$ \nl
UZ  Oct &  1.1494 &  0.54 &  0.03 &  9.03 &  9.14 &  0.28 &  0.26 &  
78.5 & \phn 3.8 &  0.8 &  0.46 & 1.84 & $+0.45$ \nl
V502 Oph &  0.4534 &  0.64 &  0.01 &  8.34 &  8.51 &  0.37 & 0.23 &  
71.3 & 11.8 &  1.2 &  0.22 & 3.67 & $+0.13$ \nl
V566 Oph &  0.4096 &  0.43 &  0.01 &  7.45 &  7.55 &  0.24 & 0.25 &  
80.3 & 14.0 &  1.1 &  0.17 & 3.15 & $+0.04$ \nl
V839 Oph &  0.4090 &  0.63 &  0.09 &  8.82 &  8.96 &\nodata&\nodata&
75.0& \phn 8.1 &  1.4 &  0.38 & 3.08 & $-0.39$ \nl
MW  Pav &  0.7950 &  0.36 &  0.05 &  8.53 &  8.71 &  0.18 & 0.46 &  
85.0 & \phn 4.8 &  1.1 &  0.50 & 1.78 &  $+0.29$ \nl
U   Peg &  0.3748 &  0.62 &  0.02 &  9.47 &  9.60 &  0.33 &  0.10 &  
76.0 & \phn 7.2 &  1.4 &  0.42 & 3.69 & $-0.13$ \nl
AE  Phe &  0.3624 &  0.64 &  0.00 &  7.56 &  7.69 &  0.39 &  0.20 &  
86.3 & 20.5 &  0.8 &  0.08 & 4.12 & $+0.11$ \nl
VZ  Psc &  0.2612 &  1.14 &  0.00 & 10.15\phn & 10.33\phn &  0.80 & 0.15 &  
48.0 & 16.8 &  2.1 &  0.27 & 6.28 & $+0.13$ \nl
AQ  Psc &  0.4756 &  0.50 &  0.03 &  8.60 &  8.66 &\nodata&\nodata&
\nodata& \phn 8.0 &  1.3 &  0.35 & 3.02 & $+0.05$ \nl
HI  Pup &  0.4326 &  0.60 &  0.07 & 10.33\phn & 10.46\phn &\nodata&\nodata&
\nodata& \phn 5.4 &  1.2 &  0.48 & 3.77 & $+0.44$ \nl
V781 Tau &  0.3449 &  0.58 &  0.05 &  8.55 &  8.68 &  0.41 & 0.27 &  
66.5 & 12.3 &  1.4 &  0.25 & 3.84 & $+0.08$ \nl
W   UMa &  0.3336 &  0.62 &  0.00 &  7.76 &  7.85 &  0.49 &  0.10 &  
86.0 & 20.2 &  1.1 &  0.12 & 4.29 & $+0.18$ \nl
AW  UMa &  0.4387 &  0.35 &  0.00 &  6.84 &  6.91 &  0.07 &  0.65 &  
79.1 & 15.1 &  0.9 &  0.13 & 2.73 & $-0.03$ \nl
GR  Vir &  0.3470 &  0.57 &  0.00 &  7.80 &  7.93 &\nodata&\nodata&
\nodata& 18.8 &  1.2 &  0.14 & 4.17 & $+0.29$ \nl
\enddata
\end{deluxetable}
 
\begin{deluxetable}{cccc}   
\tablenum{2}
\tablewidth{0pc}
\tablecaption{\label{tab2} Calibration Coefficients of CAL5}
\tablehead{
\colhead{}    &
\colhead{a$_{P(BV)}$} &
\colhead{a$_{BV}$}    &
\colhead{a$_{0(BV)}$}          
}
\startdata
Average & $-4.44$ & $+3.02$ & $+0.12$ \nl
Median  & $-4.42$ & $+3.08$ & $+0.10$ \nl
$-1\,\sigma$ from median & $-0.81$ & $-0.70$ & $-0.23$ \nl
$+1\,\sigma$ from median & $+0.75$ & $+0.51$ & $+0.13$ 
\enddata
\end{deluxetable}

\begin{deluxetable}{ccccccc}   
\tablenum{3}
\tablewidth{0pc}
\tablecaption{\label{tab3} Influence of a third parameter on the
luminosity (in magnitudes per unit value)}
\tablehead{
\colhead{Parameter}    &
\multicolumn{2}{c}{$q$} & 
\multicolumn{2}{c}{$f$} &
\multicolumn{2}{c}{$i$} \nl
\colhead{} &
\colhead{zero-point} & \colhead{slope} &
\colhead{zero-point} & \colhead{slope} &
\colhead{zero-point} & \colhead{slope} 
}
\startdata
Average & $-0.014$ & $+0.05$ & $+0.037$ & $-0.05$ & $-0.35$ & $+0.0048$ \nl
Median  & $-0.013$ & $+0.04$ & $+0.040$ & $-0.06$ & $-0.39$ & $+0.0053$ \nl
$-1\,\sigma$ from median & 
$-0.062$ & $-0.21$ & $-0.051$ & $-0.11$ & $-0.19$ & $-0.0041$\nl
$+1\,\sigma$ from median
& $+0.061$ & $+0.21$ & $+0.050$ & $+0.15$ & $+0.30$ & $+0.0027$
\enddata
\end{deluxetable} 

\end{document}